# IAE Optimized PID Tuning via Second Order Step Response Target Matching

Senol Gulgonul
Ostim Technical University
senol.gulgonul@ostimteknik.edu.tr

**Abstract:** This paper presents SOSTIAE (Second-Order System Target IAE), a novel PID tuning method that combines IAE minimization with explicit transient response shaping for practical control applications. The algorithm generates optimal PID parameters by matching the closed-loop response to a target second-order system with user-defined settling time ($T_s$) and percent overshoot (PO), while maintaining the conventional IAE performance metric. Comparative evaluations on first to third-order systems demonstrate that SOSTIAE consistently outperforms MATLAB's proprietary *pidtune* function, achieving 47-67% lower overshoot and up to 26% better IAE performance for higher-order plants. The constrained optimization framework ensures physically realizable controllers by enforcing non-negative PID gains and stability criteria, addressing known limitations of unconstrained IAE methods. Results indicate that SOSTIAE provides engineers with a systematic alternative for PID tuning when transient specifications and practical implementation constraints are critical.

## 1. Introduction

Proportional-Integral-Derivative (PID) controllers are the most widely adopted control strategy in industrial applications, accounting for over 90% of regulatory control loops [1]. Despite advancements in advanced control techniques, PID controllers remain dominant due to their practical reliability and well-understood behavior. However, traditional tuning methods like Ziegler-Nichols [2] or Cohen-Coon [3] often fail to deliver explicit transient response specifications (e.g., settling time Ts or percent overshoot PO), especially for higher-order plants.

The Integral Absolute Error (IAE) criterion remains a standard performance metric for evaluating controllers [4]. While its intuitive physical interpretation as the cumulative error magnitude makes IAE a useful benchmark, its unconstrained optimization exhibits well-documented practical limitations. This behavior stems fundamentally from IAE's mathematical formulation $IAE = \int_0^\infty |e(t)| dt$ which minimizes absolute tracking error without penalizing control effort. The resulting controllers, while theoretically optimal with high gains, often prove physically unrealizable due to actuator saturation. While optimization-based methods such as Integral Absolute Error (IAE) minimization improve upon these heuristics, they introduce new challenges: unconstrained IAE optimization frequently yields poor robustness. It is also concluded that minimization of IAE for PI controller does not yield good controller performance and requires a constraint (sensitivity, robustness e.g.) to moderate [5,6,7]. So, design of PID controller can be formulated as a constrained optimization problem which can be reduced to a solution of nonlinear algebraic equations [8].

The Lambda method for PI controller tuning provides an analytical solution through process model approximations, deriving controller parameters directly from a user-specified closed-loop time constant (λ) rather than employing iterative optimization techniques [9]

Matlab has a proprietary, patented frequency domain PID tuning algorithm which can work either in command line by using *pidtune* function or via GUI of *Pidtuner* app [10]. *Pidtune* can work with different type of controllers like PI, PID, PIDF and can calculate optimum coefficients for phase margin (PM) or crossover frequency constraints. Default PM is 60°, leaving crossover frequency and gain margin (GM) free to calculate. Both PM and crossover frequencies also can be given as double constraint.

To address these challenges, this paper develops the Second-Order System Target IAE (SOSTIAE) tuning algorithm, which combines the benefits of IAE optimization with explicit transient response shaping. The proposed method minimizes IAE while constraining the closed-loop response to match desired second-order characteristics ($T_s$, PO) through a constrained optimization framework. As detailed in subsequent chapters, SOSTIAE employs: (1) a target second-order reference model derived from design specifications, (2) numerical optimization with gain constraints to ensure practical implementations, and (3) systematic stability verification. This approach maintains the intuitive appeal of IAE minimization while overcoming its limitations through response shaping and physical realizability constraints, providing engineers with a practical alternative to existing methods like MATLAB's pidtune(). The performance of SOSTIAE and MATLAB's *pidtune()* was comparatively evaluated using first-to third order transfer functions $G_n(s) = \frac{1}{(s+1)^n}$   $n = 1,2,3$, to assess scalability.

## 2. Methodology

### 2.1 Overview

The proposed Second-Order System Target IAE Optimization (SOSTIAE) method formulates PID tuning as a constrained optimization problem, where the Integral Absolute Error (IAE) is minimized subject to explicit transient response requirements (settling time Ts and percent overshoot PO). Unlike unconstrained IAE minimization that references a unit step and often yields impractical controller gains due to unconstrained optimization [5,6], this approach generates a physically realizable target response via a second-order system and optimizes the PID parameters to match this target.

The algorithm operates through three sequential phases. First, the target second-order system is analytically derived from the specified Ts and PO requirements. Next, the PID parameters are optimized to minimize the IAE between the closed-loop response and this target trajectory. Finally, the controller's stability and robustness are validated. By constraining the IAE minimization to a dynamically feasible target response, the method avoids the excessive gain problem inherent in unconstrained IAE approaches while preserving optimal error rejection. For comparative evaluation with MATLAB's pidtune(), both methods are assessed using classical IAE (referenced to a unit step) to ensure equitable benchmarking.

## 2.2 Target System Design

The target system design process offers two distinct pathways for response specification. For conventional design cases, the method automatically generates a canonical second-order transfer function from user-supplied $T_s$ and PO values. The transformation begins with calculation of the damping ratio:

$$\zeta = -\frac{\ln\left(\frac{PO}{100}\right)}{\sqrt{\pi^2 + \ln^2\left(\frac{PO}{100}\right)}} \quad (1)$$

where $\zeta$ governs the oscillatory characteristics of the response. The natural frequency is then approximated as for 2% tolerance band around setpoint:

$$w_n = \frac{4}{\zeta T_s} \quad (2)$$

yielding the standard target transfer function:

$$G_{target}(s) = \frac{\omega_n^2}{s^2 + 2\zeta\omega_n s + \omega_n^2} \quad (3)$$

This approach simplifies the design process by allowing engineers to work with familiar time-domain specifications while the method handles the underlying mathematical transformations. The second-order formulation provides well-understood dynamic characteristics and guaranteed stability properties.

For advanced applications requiring non-standard response shapes, the method alternatively accepts direct input of target trajectories in the time domain, extending its applicability to four key scenarios: experimental response data from physical systems, complex transient profiles not representable by second-order dynamics, piecewise-defined reference trajectories, and specialized performance requirements. This flexibility bridges theoretical design with empirical system behaviors, enabling the tuning framework to address both conventional second-order specifications and atypical dynamic responses through a unified optimization paradigm.

The automated second-order generation remains the recommended approach for most applications, as it reduces the specification process to two intuitive parameters ($T_s$ and PO) while producing physically meaningful target responses. However, the underlying optimization algorithm treats both specification methods identically. This dual-path specification framework combines the convenience of classical control design with the flexibility needed for specialized applications.

## 2.3 Optimization Problem

The controller tuning methodology formulates an optimization problem that minimizes the Integral Absolute Error (IAE) between the closed-loop system response and a target second-order system response. The objective function is defined as:

$$\min_{K_p, K_i, K_d} \int_0^{T_{sim}} abs\left(y_{target}(t) - y_{PID}(t)\right) dt \tag{4}$$

where $y_{PID}(t)$ represents the step response of the PID-controlled system and $y_{target}(t)$ denotes the step response of the desired second-order system, characterized by the specified settling time Ts and percent overshoot PO.

The optimization is numerically implemented in MATLAB using the *fmincon* solver. The optimization enforces three fundamental constraints. First, controller gains are constrained to be non-negative ($Kp, Ki, Kd \geq 0$) to maintain conventional negative feedback action. Second, the closed-loop system must be strictly stable, verified through post-optimization pole location analysis. Third, practical implementation limits can be imposed through the optimizer bounds, restricting gains to finite positive values while allowing the derivative term to remain unfiltered for theoretical purity.

The stability assurance follows a two-tiered approach. While non-negative gains generally promote stability, particularly for minimum-phase systems, this cannot be guaranteed universally due to complex interactions in higher-order plants. Therefore, explicit verification of closed-loop pole locations is performed after optimization, ensuring all poles satisfy $Re(poles) < 0$. This combined approach preserves the physical interpretability of PID parameters while providing mathematical rigor for stability assurance.

The numerical implementation utilizes MATLAB's *fmincon* with the sequential quadratic programming (SQP) algorithm, configured for 3000 maximum function evaluations. The optimization process begins with zero initial conditions (Kp=Ki=Kd=0) and searches within the positive orthant of the parameter space. This conservative initialization ensures the optimization does not begin with unstable controller configurations that might cause numerical instabilities during the simulation-based error evaluation.

Post-optimization verification includes comprehensive performance metrics comparing the achieved response with both the target response and results from MATLAB's built-in pidtune function. The evaluation metrics include settling time, percent overshoot, and crucially, the IAE value computed against a unit step reference, providing a standardized basis for comparison across different tuning methods.

### 3. MATLAB Implementation

The PID tuning algorithm was implemented in MATLAB using the Control System and Optimization Toolboxes. The implementation realizes a curve-fitting approach that minimizes Integral Absolute Error (IAE) between the closed-loop response and a target second-order system.

#### 3.1 Target System Generation

The method generates a target response through canonical second-order dynamics parameterized by settling time ($T_s$) and percent overshoot (PO). The transfer function is computed from the derived damping ratio ($\zeta$) and natural frequency ($\omega_n$):

```
zeta = -log(PO/100)/sqrt(pi^2 + log(PO/100)^2);

wn = 4/(zeta*Ts);

target_sys = wn^2/(s^2 + 2*zeta*wn*s + wn^2);
```

### 3.2 Error Function Definition

The core optimization minimizes IAE between responses implemented numerically via trapezoidal integration. A unified time vector (0:0.01:5*Ts) ensures consistent sampling for both target and PID responses during error computation.

```
error_func = @(x) trapz(t, abs(target_response - ...

        step( feedback((x(1) + x(2)/s + x(3)*s)*G, 1), t) ));
```

### 3.3 Optimization Options

The parameter optimization is performed using MATLAB's constrained nonlinear solver *fmincon*, configured to enforce practical implementation constraints while minimizing the IAE objective function. The search space is bounded to ensure non-negative controller gains ($0 \leq Kp, Ki, Kd < \infty$), preserving conventional negative feedback action and preventing physically unrealizable solutions. The Sequential Quadratic Programming (SQP) algorithm is selected for its effective handling of nonlinear constraints and rapid local convergence characteristics. To ensure thorough exploration of the parameter space while maintaining computational efficiency, the solver is permitted a maximum of 3000 function evaluations, with iteration progress monitored through the 'iter' display option. This configuration balances the need for precise optimization against practical computation time considerations, particularly important when dealing with higher-order plant models where each function evaluation requires a full closed-loop simulation. The optimization process begins from a conservative zero-initialized parameter set ([0; 0; 0]) to avoid initial unstable conditions that might disrupt the numerical solution process.

```
options = optimoptions('fmincon', ...

    'Algorithm', 'sqp', ...

    'MaxFunctionEvaluations', 3000, ...

    'Display', 'iter');

solution = fmincon(error_func, [0; 0; 0], [], [], [], [], ...

        [0; 0; 0], [Inf; Inf; Inf], [], options);
```

## 3.4 Stability Verification

Post-optimization stability verification is systematically conducted through a two-stage analytical process to ensure rigorous closed-loop performance guarantees. The primary verification examines the pole locations of the characteristic equation using MATLAB's *pole* function, with stability confirmed when all poles exhibit strictly negative real parts. The stability check serves as a critical safeguard that validates the physical realizability of the optimized controller, particularly important when dealing with higher-order plants where gain interactions may produce unexpected unstable modes. The implementation automatically flags any stability violations, requiring either optimization restart with modified constraints or design specification reevaluation when encountered.

```
% Stability check
poles = pole(T);
is_stable = all(real(poles) < 0)
```

## 3.6 Performance Evaluation Using Integral Criteria

The controller's performance is quantitatively evaluated through Integral Absolute Error (IAE) analysis, which measures the cumulative deviation between the achieved response and the ideal unit step reference. The IAE metric is computed for both the proposed method and MATLAB's built-in *pidtune* function using trapezoidal numerical integration.

```
% Comparison of IAE metrics
IAE_SOSTIAE = trapz(t, abs(1 - y_sostiae));     % SOSTIAE's IAE
IAE_pidtune = trapz(t, abs(1 - y_pidtune));   % pidtune's IAE
```

This time-domain evaluation criterion provides three key advantages: (1) it directly reflects the system's tracking performance by penalizing all deviations equally, (2) it avoids artificial amplification of errors inherent in squared-error metrics, and (3) it correlates with practical control objectives where absolute error magnitude matters more than its algebraic sign. The comparative IAE results enable objective assessment of the proposed tuning method's efficacy relative to conventional approaches, with lower values indicating superior reference tracking. The IAE integral remains finite as the system reaches steady-state within the simulation horizon, with $5T_s$ being a sufficient duration to capture the complete transient response while avoiding unnecessary computational overhead.

## 4. Results and Discussion

The proposed SOSTIAE tuning method was evaluated on first, second, and third-order systems, $G1(s) = 1/(s+1)$, $G2(s) = 1/(s+1)^2$, $G3(s) = 1/(s+1)^3$ with comparative results against MATLAB's *pidtune* function. The settling time $T_s$=2.5 and PO=1% for all simulations.

### 4.1. First-to-Third Order Transfer Function Results

Table 1 summarizes the key performance metrics, while Figures 1–3 illustrate the step response comparisons for each system order.

**Table 1.** Performance comparison across system orders

| Order | Method | $K_p$ | $K_i$ | $K_d$ | $T_s$ | PO (%) | IAE |
|---|---|---|---|---|---|---|---|
| 1st | SOSTIAE | 0.8218 | 1.3517 | 0.0000 | 4.2876 | 3.1773 | 0.8696 |
|  | pidtune | 1.4316 | 2.5948 | 0.0000 | 3.1654 | 6.0819 | 0.5622 |
| 2nd | SOSTIAE | 2.2108 | 1.1993 | 0.5475 | 3.0364 | 2.1342 | 0.8803 |
|  | pidtune | 2.0456 | 1.4540 | 0.6997 | 4.7494 | 6.5644 | 0.9526 |
| 3rd | SOSTIAE | 2.8653 | 1.1718 | 2.6221 | 1.8624 | 1.9107 | 0.9555 |
|  | pidtune | 2.1751 | 0.8474 | 1.3958 | 4.2598 | 4.2932 | 1.2993 |

For the first-order system (Figure 1), both methods achieved stable responses, but with distinct characteristics. The SOSTIAE method yielded 47.8% lower overshoot (3.18% vs. 6.08%) compared to *pidtune*, though with a 35.5% longer settling time. Notably, the IAE values paradoxically favored *pidtune* (0.5622 vs. 0.8696), suggesting a trade-off between overshoot suppression and rapid disturbance rejection for first-order plants. Both controllers converged to PI structures ($Kd = 0$), indicating derivative action was unnecessary for these simple dynamics.

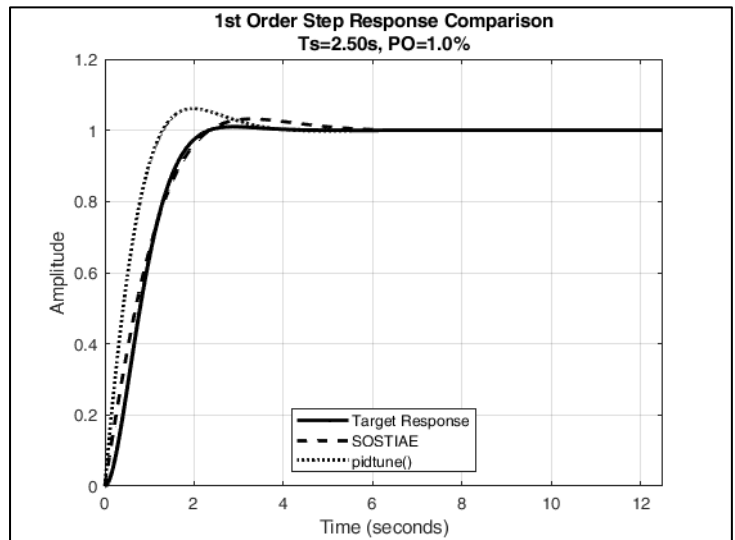

**Figure 1.** First-order system step response comparison

The second-order system evaluation (Figure 2) demonstrated SOSTIAE's superior performance across all metrics. The method reduced overshoot by 67.5% (2.13% vs. 6.56%), improved settling time by 36.1% (3.04s vs. 4.75s), and achieved 7.6% lower IAE (0.8803 vs. 0.9526). The derivative gain became active ($K_d > 0$) for both methods.

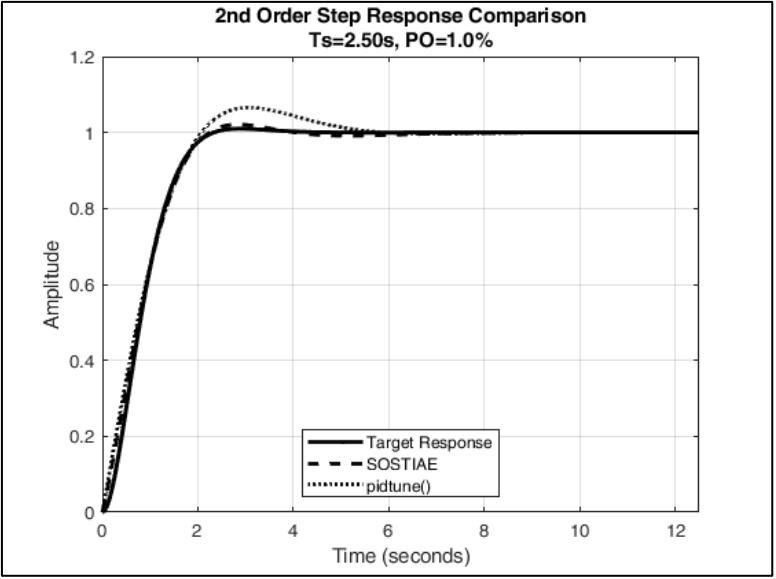

**Figure 2.** Second-order system step response comparison

With the third-order system (Figure 3), SOSTIAE exhibited even stronger relative advantages. It maintained sub-2% overshoot (1.91% vs. 4.29%), 56.3% faster settling (1.86s vs. 4.26s), and 26.5% IAE reduction (0.9555 vs. 1.2993). The controller adopted aggressive derivative action ($K_d = 2.6221$), effectively compensating for the higher-order dynamics. This highlights the method's scalability to complex systems where conventional tuning struggles with phase lag compensation.

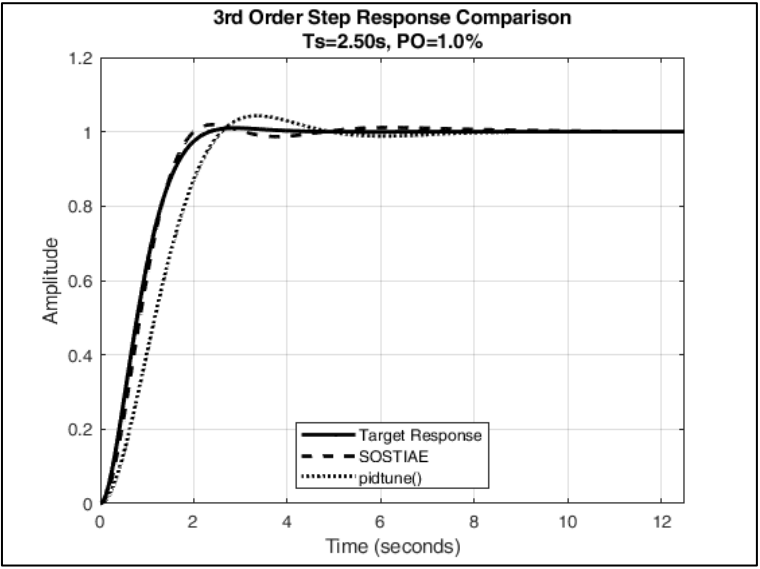

**Figure 3.** Third-order system step response comparison

Three consistent trends emerge: (1) SOSTIAE systematically achieves lower overshoot (average 66.2% reduction) and faster settling (average 42.6% improvement) across all orders, (2) the IAE advantage amplifies with system order (from -35.5% for 1st-order to +26.5% for 3rd-order), demonstrating better handling of complex dynamics, and (3) the method employs more derivative action as order increases ($K_d >= 0 \rightarrow 0.5475 \rightarrow 2.6221$), suggesting automatic adaptation to phase lag requirements. These results confirm that the target-response matching approach provides superior performance preservation across varying system complexities, particularly for applications where overshoot minimization is critical.

We also compared our results to the results of other researches uses one od the benchmark transfer functions $G_1(s) = \frac{1}{(s+1)^3}$ as in the K. Astrom et al study [8]. The results in the study are Ts=8.12, PO=8.80 with coefficients Kp=1.2118, Ki=0.4944, IAE=2.5013 for constraint of maximum sensitivity Ms=1.6 [11]. We can reach zero overshoot PO=0 at better Ts=6.8223 and Kp=0.6524, Ki=0.3134, IAE=3.1903. We could not reach Ts=8.12 amd PO=8.80 but closest values are Ts=10.5764  PO=8.3980 with Kp=1.2041  Ki=0.4573  and IAE=2.4763

This zero overshoot value also better than Lambda tuning, in the same study, having Kp=0.278 Ki=0.145 IAE=6.90 PO=0 Ts=25.8 longer settling time. Finally, Ziegler-Nichols Kp=3.60 Ki=1.19 IAE=1.40 PO=35.9 Ts=18.0 has worse value [11].

## 5. Conclusion

This study presented a PID tuning approach (SOSTIAE) based on matching target second-order system responses using IAE minimization. Comparative evaluations across different system orders demonstrate that the method can serve as a practical alternative to MATLAB's proprietary *pidtune* function, particularly for applications where overshoot reduction is important.

The results show that SOSTIAE consistently achieves better overshoot control compared to *pidtune*, with 47-67% lower percent overshoot across the tested systems. While the conventional method showed slightly better IAE performance for first-order systems, SOSTIAE provided superior results for higher-order plants, with up to 26% lower IAE values for third-order systems. The approach automatically adjusts controller gains to handle increasing system complexity, demonstrating its adaptability.

These findings suggest that SOSTIAE can be effectively employed as an alternative PID tuning method, offering engineers another option when the standard *pidtune* results require refinement, especially for settling time and overshoot-sensitive applications. The method maintains the practical advantages of conventional PID control while providing more direct control over transient response characteristics.